# Physical Memory Attacks and a Memory Safe Management System for Memory Defense

Alon Hillel-Tuch[1] [0000-0002-2307-3486] and Aspen Olmsted[2][0000-0003-2652-0154]

[1] New York University Tandon School of Engineering, Brooklyn, NY 11201, USA
[2] Simmons University, Boston, MA 02115, USA
ah5647@nyu.edu olmsteda@simmons.edu

**Abstract.** Programming errors, defective hardware components (such as hard disk spindle defects), and environmental hazards can lead to invalid memory operations. In addition, less predictable forms of environmental stress such as radiation, thermal influence, and energy fluctuations can induce hardware faults. Sometimes, a soft error can occur instead of a complete failure, such as a bit-flip. The 'natural' factors that can cause bit-flips are replicable through targeted attacks that result in significant compromises, including full privileged system access. Existing physical defense solutions have consistently been circumvented shortly after deployment. We will explore the concept of a novel software-based low-level layer that can protect vulnerable memory targeted by physical attack vectors related to bit-flip vulnerabilities.

## 1    Introduction

Natural memory faults can manifest with no detectable hardware defect, be temporal, and not predictably replicable. Recovering from these errors at the Operating System (OS) level is non-trivial. However, once personal computing machine ownership became ubiquitous, researchers developed hardware error correction solutions for these naturally occurring faults that remain foundational in mainstream hardware-based Error Correction Controllers (ECC) memory applications [1]. Hardware component sophistication may vary significantly depending on the operating requirements of the commercial, industrial, and residential environments, blurring the lines of personal computing.

ECC is a mature hardware solution to detect and potentially correct natural memory faults. However, consumer-grade memory up until and including DDR4 predominantly did not contain ECC - certain consumer DDR4 does have ECC but limited support from Central Processing Unit manufacturers. The JEDEC standard for DDR5 has made an ECC variant part of the standard [2]. However, this would be an on-die solution that detects on-die errors, such as bit-flipping, and does not protect data in transit between the on-die memory controller and the central processing unit; it is not as comprehensive as a DIMM-wide ECC solution.

Furthermore, when facing sophisticated attacks that exploit physical hardware attributes such as RowHammer [3], ECC provides limited protection and, at best, will slow down the attack, as demonstrated in the ECCPLOIT RowHammer-like attack that can induce bitflips in server-grade ECC memory [4]. Server equipment typically features a more sophisticated memory controller with hard and soft protections embedded in their firmware. Consumer hardware does not have any advanced solutions, with limited adoption of ECC functionality, and relies on software-based solutions, if at all. Research exploring the balance between the overhead costs from software-based error detection, and expected gains from data integrity, has implicitly highlighted that the balance is subjective to the computational task [5]. We posture that consumers have a low tolerance for increased computational difficulty (latency) and propose a novel solution that will provide the computational work only for critical data regions, which are a relatively small component of all computational tasks performed.

The organization of the paper is as follows. In section 2, we discuss our motivating example. Section 3 describes the related work and the limitations of current methods. Section 4 describes our hypothetical solution and experiment parameters. Section 6 drills into the data gathered in our experiment. Section 6 expands on the hypothesis to develop an implementation concept. Finally, we conclude in Section 7.

## 2      Motivating Example

We postulate that memory experiencing randomized hardware faults with no-fault detection by the system will remain prevalent in the foreseeable future due to legacy hardware and solutions in the consumer space. Research has explored self-healing OSs that minimize fault-induced downtime and halts [6] [7]. In addition, the transition from pure monolithic kernel structures has improved robustness. However, there is little OS consideration around memory validity beyond in-hardware by the memory manufacturer. Furthermore, specific faults can result in data that remains syntactically valid. Without a fault condition signaled on the CPU bus, there is no interrupt and no OS intervention, potentially resulting in unexpected behavior. This behavior can compromise system security popularized by case studies like Dinaburg's bitsquatting research [8].

Beyond bitsquatting, targeted attacks can induce memory errors in specific memory regions. Certain kernels are susceptible to these disturbance-based error attacks as they use kernel-same-page merging (KSM) to optimize memory page usage. The process by which most deduplicate memory is to scan the entire main memory and identify physical pages that hold identical page content (duplicates). After identifying the target pages, we can look at the virtual pages that point to those physical pages. Instead of removing the duplicate virtual page, we modify the pointers of one of the virtual pages to point to the same physical page as the other virtual page, allowing us to remove one of the two physical pages (or more if there are >2 duplicates).

Within specific kernels, such as Linux, Copy-On-Write is enabled to make sure that if such a process (shared virtual page) does end up modifying the physical page, a copy will be made on write to avoid conflicts among the virtual pages. However, we can instantiate a virtual page containing content that matches the type of page that would contain access controls, signatures, and other similar data. We then have a page that can resolve to point to the same physical page in memory, allowing us (from a contained environment) to apply targeted attacks that may induce bit-flips at high precision.

Flip Feng Shui is a RowHammer attack technique that can operate from within a Linux VM and successfully compromise DDR3 and DDR4 memory. Researchers have used Flip Feng Shui to break OpenSSH public-key authentication and spoof GPG signatures from trusted keys [15]. Research has also demonstrated new methods such as 'Blacksmith' that circumvent Row Refresh (TRR) mitigations implemented on physical memory in response to RowHammer [9].

RowHammer and Blacksmith attacks require minimal disk space and no network access for the attack itself. However, they require CPU time and memory resources due to memory's significant 'hammering.' OSs with native Power Management tools that include active monitoring may flag the process as memory intensive. Furthermore, running the attack within a virtual machine may circumvent heuristic scanners and anomaly detectors due to the behavior being 'normal' for the process type, such as the vmmem process in WSL2, VMware Workstation, or specific Docker containers. Considering cloud hosting services rely heavily on virtualizations of Linux, this is a genuine concern.

The primary motivating example in our research is that most hardware-based protections are highly specific. On the other hand, circumvention is relatively simple once identified and often based on pattern detection, which is maskable. Current soft solutions can provide updatable comprehensive memory protection strategies. However, while compelling, soft solutions are complex and, at present, do not differentiate between critical and non-critical data, causing increased computational complexity due to broad-based redundant multithreading (or single) and space complexity [10-12].

## 3    Related Research

### 3.1    Research Fields

Research into software-based memory error detection emphasizes creating safe access solutions, such as MEDS and memory-safe languages, versus addressing hardware faults [16]. Hiser et al. focused on programming languages, and the prevention of memory overwrites (a common form of memory error in non-memory safe languages), providing a solution and intermediary protection when only compiled binaries exist. However, MEDS does not protect 'seemingly' valid data, which can occur in a bit-flip from either natural or manipulated causes, nor does it genuinely acknowledge the hardware level, mainly remaining in the virtualized memory address space. Our research shares common challenges the MEDS team faced in handling low-overhead translation of their monitor.

Savaria and Velazco developed a comprehensive suite of error detection mechanisms that can complement and, in some cases, replace hardware solutions [17]. Their answer toward complete bit-flip detection is highly applicable to industrial and memory-critical deployments. The computation and space cost of an instruction-duplication-based or signature-based technique is generally acceptable in these data-critical situations. However, in consumer environments, the overhead may not be considered tolerable when contrasted to the relatively high degree of non-critical data used in computations and the expectation of low latency (responsiveness). The solution results in a "speed decrease by a threefold factor and required memory is four times larger" (p. 3517).

Additional work from independent researchers and cybersecurity firms around DinasBurg's bitsquatting concept has demonstrated that consumer-grade hardware is susceptible to the repetitive nature of specific tasks and will treat Words containing bitflips, due to the lack of error detection, as valid [18-19]. Hax identified a core example within the Networking Time Protocol (NTP). Windows synchronizes to the NTP servers regularly, creating many trials. However, the expected value is still significant even with the low probability of a naturally induced bit-flip. Allowing for payload attacks such as the 2038 problem, an exploit causing an overflow in the signed 32-bit integer used to store the time on older OSs [20].

The Blacksmith research successfully induced bit-flips in all 40 DRAM devices selected from the three most prominent vendors supplying memory cells and controllers. Unfortunately, the defense mechanisms manufacturers implemented against RowHammer-like attacks are proprietary and unknown to the public. That said, the research from Jattke et al. obtained a high success rate circumventing the protections (CVE-2021-42114). We see this as a clear demonstration of the binary nature of these hard defenses, initially solid and practical, but once usurped, they become futile and are typically hard to patch, if at all, due to their embedded nature. Furthermore, research has demonstrated an ability to "factor 4.2% of the two 4096 bit Ubuntu Archive Automatic Signing keys with a bit-flip", allowing the researchers to induce a victim VM to install pre-determined Linux packages for which the install privileges should not exist [15].

The Ubuntu signing key attack is part of a broader exploration in Razavi et al.'s discussion of software mitigation techniques for their exploitation vector Flip Feng Shui (FFS), which uses RowHammer and memory deduplication [15]. Most of the soft solutions discussion is focused on the disabling of memory deduplication and performing an analysis of the various methods by which to as efficiently as possible avoid data bloat due to the lack of identical page merging. There is mention of how "security-sensitive information needs to be checked for integrity in software right before use to ensure the window of corruption is small." [p.15]. The suggestions made pertain to the type of data stored and its format, for example, the X.509 certificate chain format. However, the issue remains an integrity and authentication concern at the local level, including the filesystem and the metal level between kernel and hardware. A successfully induced bit-flip can result in the ability to usurp access controls, bypass sandboxes, and gain kernel-level privileges [21]. We must assume that the operating and file systems cannot be trusted.

## 4   Hypothesis & Parameters

We classify software operations into two prioritizations, critical and non-critical (priority flags). Non-critical operations address memory which can remain susceptible to bitflips. Their integrity is not essential for system security or the general user experience. A priority flagged operation will enjoy an enhanced integrity process with the reference monitor coordinating the execution of additional data validity algorithms to ensure data integrity and providing restrictive access controls to relevant memory regions. While those additional operations will consume extra time and space complexity discussed in sections 2 and 3, they are a small subset of general software operations. The anticipation is that the OS will have a minimal perceived impact increase in process latency. We seek to provide a more balanced solution for randomized hardware faults by allowing selectivity when operating computationally demanding memory integrity solutions. Our solution will reduce OS hard-faults and enhance system security from RowHammer, Blacksmith, and other disturbance-based error attacks [13-14].

Furthermore, we believe that we must treat the filesystem as a compromised component when acknowledging that the OS itself cannot be trusted. For example, the existence of a page cache is exploitable. While the scheduler concept mentioned above provides a conceptual approach to data verification and integrity, we believe a clear delineation of memory is needed. Nider and Rapoport have campaigned for partially isolating the Linux Kernel alongside areas of responsibility [23]. We would consider the priority flag and any data needed by the reference monitor and any selected algorithm to provide data validity to be protected and assigned a single owning concept control accessible only via the reference monitor following NEAT [24]. We propose this two-part system as the *Memory Safe Management System (MSMS).*

To demonstrate the concept of data prioritization, we will conduct a simulation that artificially injects hardware faults and will measure detection, space complexity, and time complexity. For our current demonstration, we will use an overly simplified parity-check algorithm. It is not an appropriate data validity algorithm in any real-world environment. Our goal in this initial research is to outline the concept and demonstrate high-level viability. The main benefit of a software solution is to select a preferred memory integrity algorithm for the task at hand. A modular construct will permit this.

We can create computed Berger code check-bits for high priority words using a modified Berger code system for 10110, {{bit zero error, {**0**0110,10**0**10,10100}}, {bit one error, {1**1**110,1011**1**}}. We store the parity of a Word as a single bit, and upon retrieval, we recalculate the parity and compare it to the value stored in the bit. The priority-bit is OS-directed, and the expectation is that programs will declare which operations are priority-based; the scheduler will then assign services.

We then simulate random 8-bit memory operations (n = 4,729,000), a randomized single-bit soft-error rate, and a randomized priority classification. Our analysis will compare error detection against the increase in computational overhead between the historical method of operation and our scheduler-based parity-bit detection. We classify 15% of all operations as high priority (critical). No Detection means no active mechanism, Enhanced Detection means we only apply our strategy on priority classified operations, and Full Detection means we use our approach on all memory operations.

To introduce an error probability, we leverage research from Intel which initially discovered that alpha-particle bombardment is a measurable inducer of soft errors in DRAM [22]. IBM's study further quantified the observation into a probabilistic model that considered factors beyond just alpha particles, allowing us to define the probability of a soft error [22]. As a result, 1GB SDR ECC Memory is expected to express 900 fails per 10,000 systems per three years, estimated at one mistake per month for every 256MB of ram. Using this information, we assign a rate for total single-bit errors introduced artificially in the simulation: $Error\ Probability\ Factor = 1.6 \times 10^{-5}$, $expected\ max\ error = 7.5 \pm 1.5$).

These factors generate a random probability applied to every simulated operation. We store a single parity bit for each 8-bit variable held in an independent C Struct; a production implementation would be native to the OS, adopting more sophisticated security considerations since the parity bit itself would be susceptible to manipulation. We capture this information alongside potential detection whenever an artificially induced error is planned. Our simulation stores the results in a CSV for analysis. We expect the system to identify all errors in priority flagged operations.

## 5 Empirical Evidence

Figure 1 illustrates the overhead variation between the three different strategies. The computational overhead for each strategy (measured as a unit of simulated steps) should all follow expected differentials due to the simplicity of the simulation. To calculate parity, we must traverse the full n-length of the Word Θ(n). The additional overhead of Enhanced Detection follows the equation S+(P×S) with P equal to the probability of a priority designation and S equal to the number of original operations plus two (one write operation and one read operation of the priority bit). Complete detection equals twice the number of initial steps (a constant, in this case, set as the Word bit-length divided by two) plus two. Computational Complexity is Θ(|Word|). Space complexity is Θ(1) due to the storage of a single parity bit and a single priority bit regardless of the size of Word. Note that even with no detection, we still perform computational work, such as reading the Word. The space and time complexity analysis rely on an overtly simple party bit flag in our experiment. This approach is not comparable to more sophisticated production-level algorithms. We do not measure the unit of space occupied. Subsequent research is required to implement the proposed reference monitor, which we do not undertake in this research.

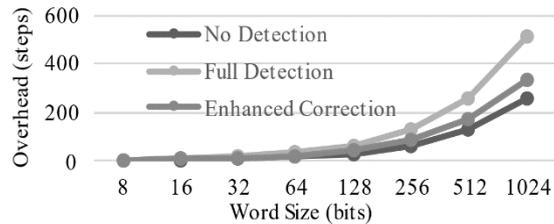

**Fig. 1.** Computational Overhead Analysis from resulted data

Figure 2 demonstrates that we detected only one error out of the eight introduced, an 87.5% miss rate. Again, a logical value, as it should follow P=.15 quite closely. The additional overhead involved in the enhanced operation follows (P ×S), directly related to the size of the Word and the percentage of Priority operations. Looking at Savaria and Velazco's Software Implemented Error Detection (SIED) solutions, we observed their discussion of a threefold factor speed decrease while increasing required memory by four [17]. We recognize that using our priority scheduler with their SIED assigned to our reference monitor can provide a basic theoretical comparison, as shown in Table 2.

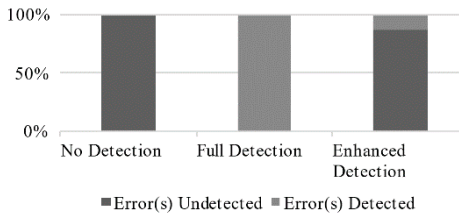

**Fig. 2.** Error Detection Analysis from resulted data

Assuming uniform data packet sizes and computational complexity for each process (an unrealistic assumption), we have 100 units of time and space. Our proposed scheduler would increase time overhead by 45% and space overhead by 60% compared to a blanket SIED implementation that increases time overhead by 200% and space overhead by 300%. We must recognize that our solution requires additional steps in assigning priority and reading priority flags. Those steps are minimal, but future research is warranted, including a solution implementation within an OS for testing.

**Table 2.** Theoretical Performance

| System | Time Units | Space Units |
|---|---|---|
| None (standard) | 100 | 100 |
| SIED | 300 | 400 |
| MSMS w/ SIED 15% priority | 145 | 160 |

The purpose of this demonstration is to show that even with the high miss rate of our solution, the seriousness of the error is substantially lower as it is a non-priority miss. By differentiating between priority and non-priority, we can selectively choose when to incur additional time and space complexity to provide data integrity. We believe this to be particularly beneficial to non-urgent background processes that use system downtime to perform tasks (such as time synchronization) and memory containing access control information. The detection mechanism will detect if memory has been altered, including attacks. We can now become aware of an integrity issue and choose an alternative action rather than move forward with the data retrieval task.

## 6 Implementation

### 6.1 Reference Monitor Priority Scheduler

To develop a realistic implementation, we need to bifurcate the scheduler (the system that assigns and reads priority flags) from the mechanism used to validate memory. Figure 3 shows one such bifurcation as a simplified process flow diagram. This initial process flow diagram does not consider a kernel's numerous memory access procedures and how to optimize for them appropriately. For example, instead of continuously retrieving the priority bit and performing a check for every single memory operation, we may choose to segment the prioritized memory entirely with a strict access path via the reference monitor. However, memory isolation provides other risks, such as easier identification and targeting by potential attacks, even virtual segmentation. For new processes (C), we would have to apply our data integrity algorithm of choice and store it appropriately. Next, we need to GET the data via the memory algorithm controller (B) for existing processes, managing data integrity. It would either return the data as valid, return and mark it as invalid, or not return and indicate invalid. These are parameters we can set. If the data is invalid and there is no error correction mechanism, the original process would need a mechanism to repeat the initial process call, discard the data, or still access the data knowing it is invalid.

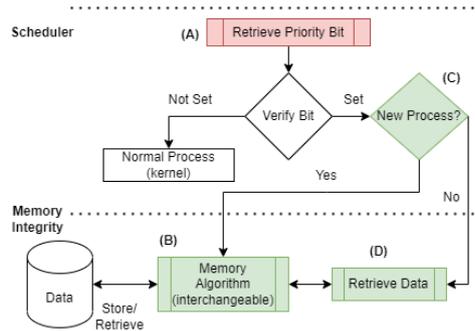

**Fig. 3.** Conceptual Process Flow for MSMS Priority Based Scheduler

### 6.2 Memory Mapping and Zones

The scheduler needs to leverage a reference monitor that operates as a memory control layer within the MSMS to provide the required memory protection. Figure 4 depicts how the scheduler from Figure 3 depends on the reference monitor to provide memory-safe access that includes secure virtual page management and priority flag-based memory assignment and protection. We can set priority within the Process Control Block or page-by-page with this kind of system. However, the scheduler will still need a global priority flag.

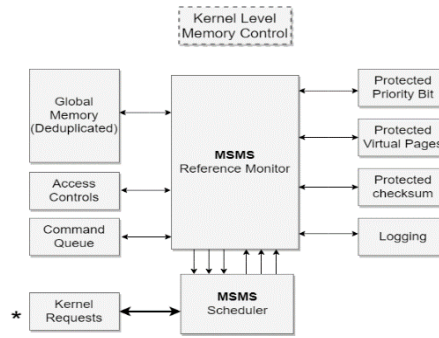

**Fig. 4.** The conceptual structure of the Memory Safe Management System (MSMS)

The right-hand side of the reference monitor indicates memory classifications that would benefit from memory isolation, with strict access controls governed by the reference monitor. The individual silos are critical to isolating each aspect of our system further. For example, the priority bit could be in a separate memory area from the protected (not deduplicated) pages. The checksum or other signature hash calculated are stored in another location. Logging is a beneficial system administrator feature to help further understand which processes are accessing what information. The logs should attempt to be incorruptible and irreparable until they have been securely offloaded from the local system. The MSMS governs access to our scheduler priority flags, manages memory isolation, and performs any necessary cryptographic functions (such as the checksum) to sign data. It would have to operate within kernel space, with the MSMS reference monitor providing a singular user-space access point. Containing the commands in a queue alongside similar queues allows us to introduce a certain degree of noisiness to the data set to ensure non-predictable checksums (or perform salting). The MSMS is only accessible by the kernel and nothing else. As a standalone environment, it can provide the necessary services. Note that the kernel cannot 'change' a priority flag. The flag is either set by the application developer or the OS process. Once the flag is implemented, we can only go from '0' to '1' but cannot change a '1' into a '0'.

## 7   Conclusion and Future Work

Implementing a scheduling system to handle flagged data validity checks appears to have merit. However, the practical implementation of a memory-safe management system depends on additional factors beyond computational and space overhead: cost, interoperability, and political will. In addition, we have yet to consider OS kernel implementation. Our proposed system would require kernel-level modifications and a discussion on good-actor behavior by developers, both challenging potential feasibility. However, an open-source operating based on the Linux kernel would enable us to construct a testing bed to develop an MSMS implementation. Another point of consideration is the security of using a singular bit to provide service, where this bit is stored, and the additional memory used to provide error detection (and potentially correction).

Furthermore, this solution does not prevent side-channel attacks. Due to the concentration of critical memory operations within one software system (the MSMS), we can argue that the ability to perform a side-channel attack could have been simplified for an attacker. Communication with the memory management system and between its components would benefit from a robust encryption solution. The ubiquity of the Trusted Platform Module and support in Windows 11 and the Linux since Kernel revision 3.2 is worth exploring to provide storage of keys and hardware acceleration of data decryption. Further exploration of memory control (including on-die microcontroller memory) and its coexistence with our proposed system is warranted.